\documentclass{ws-rv9x6}
\usepackage{subfigure}   
\usepackage{ws-rv-thm}   
\usepackage{ws-rv-van}   
\usepackage{hyperref}
\makeindex

\def\ket#1{|#1\rangle }
\def\bra#1{\langle #1 |}

\begin{document}

\chapter[Topological phases in amorphous matter]{Topological phases of amorphous matter}
\label{ra_ch12}

\author[Adolfo G. Grushin]{Adolfo G. Grushin\footnote{adolfo.grushin@neel.cnrs.fr}}

\address{Institut N\'eel, CNRS \& Universit\'e Grenoble Alpes, 38042 Grenoble, France}

\begin{abstract}
Topological phases of matter are often understood and predicted with the help of crystal symmetries, although they don't rely on them to exist. In this chapter we review how topological phases have been recently shown to emerge in amorphous systems. We summarize the properties of topological states and discuss how disposing of translational invariance has motivated the surge of new tools to characterize topological states in amorphous systems, both theoretically and experimentally. The ubiquity of amorphous systems combined with the robustness of topology has the potential to bring new fundamental understanding in our classification of phases of matter, and inspire new technological developments. 
\end{abstract}
\body

\tableofcontents

\section{Introduction}\label{sec:intro}

Topological phases of matter are among the most robust phases of nature. 
Their many-body wavefunction cannot be described with purely localized orbitals, and the information that defines these phases is stored non-locally, spread over the entire system. Therefore, their physical properties, such as their metallic boundary states or quantized responses to external fields, are protected from local perturbations, such as defects, impurities, or other material imperfections.

This chapter embraces two remarkable observations. First, it was recently predicted that around $30\%$ of all non-magnetic crystals are in a topological phase~\cite{Tang18,Zhang:2019tp,Vergniory:2019ub}, and at least a similar proportion is expected for magnetic materials~\cite{Watanabe:2018bc,Xu20,Elcoro2020}. Second, the previous chapters in this book should have convinced the reader that nearly all materials can be prepared as amorphous solids if cooled fast enough and to low enough temperatures~\cite{Turnbull69}. Here we describe the blooming research on the interface between the above two observations. It demonstrates that topological properties can survive even under extreme structural disorder, as in amorphous systems. We will describe the recent developments that have characterized theoretically and experimentally topological phases in amorphous solids. Along the way, we will mention other states that host topological phases without translational invariance, such as quasicrystals, topological Anderson insulators, and fractals. 

Amorphous and topological matter share a great technological potential~\cite{Zallen,Anonymous:2016tp}. Due to its versatility, the former is around us, in glass windows, plastics, fiber optics, and solar cells. Due to its robustness the latter has been identified to hold the key to robust quantum computation, memory storage, chiral waveguides, spintronics, metrology, and thermo-electricity. Hence, it is likely that by combining these two forms of matter we can develop a broader understanding and classification of the properties of solids, and a richer set of applications than those separately devised for the two fields.

Since this book focuses on amorphous matter and glasses and not topological states, we feel compelled to introduce the basics of topological matter to the reader. We will not give rigorous proofs, but rather examples and plausibility arguments that illustrate the main properties associated to topological states. This introduction will contextualize the rupture that amorphous topological matter signifies for our current understanding of topological phases, especially due to the central role that symmetry plays in determining topological properties.

Before we commence, a word of warning. In this chapter we will use the term topological phases to mean non-interacting topological phases. Interacting topological phases, such as fractional quantum Hall states, are usually referred to as \emph{topologically ordered} states\cite{Wen17}. The extension of the developments presented in this chapter to interacting phases remains an open question, to which we come back upon concluding.

\section{A primer on topological matter: the role of symmetry}

The concept of symmetry is deeply rooted in our understanding of condensed matter, and topological insulators and metals are not an exception. Although topological phases can exist in the absence of any symmetry, symmetry enriches their classification, and allows to express the topological invariants that define these phases and their responses in mathematically simple ways. As we will see, the concept of translation symmetry, although strictly unnecessary, is central to our understanding.

The basic property of an insulator is the absence of electrically conductive states. The simplest of all insulators is the atomic insulator, a set of disconnected atoms with electrons forming closed shells. If these atoms form a crystal lattice, we may use Bloch's theorem to combine these shells into energy bands separated by finite gaps. Since electrons are bound to the atoms they lack kinetic energy, and the bands $E_n(\mathbf{k})$ are exactly flat as a function of crystal momentum $\mathbf{k}$. Now imagine slowly bringing these atoms together, forming covalent bonds in the process. The electronic bands may now disperse with $\mathbf{k}$, and the gap sizes can change. If at all stages in the process we retained the existence of a band gap we may say that the two insulators, the atomic limit and the covalently bonded solid, are smoothly connected to each other by varying a parameter, in this case the strength of the covalent bonding. We call the insulators connected to the atomic insulator in this way topologically trivial insulators.
 
Topological insulators present fundamental obstructions to reach the atomic limit. The smooth process that connects them to the atomic insulator will always lead to a gap closing, and with it, fundamentally different responses. This defines topological insulators as insulators that cannot be smoothly transformed into an atomic limit. In many instances this obstruction is due to an underlying symmetry; there is no path that simultaneously keeps the gap open and respects the symmetry.

\begin{figure}
    \centering
    \includegraphics[width=\textwidth]{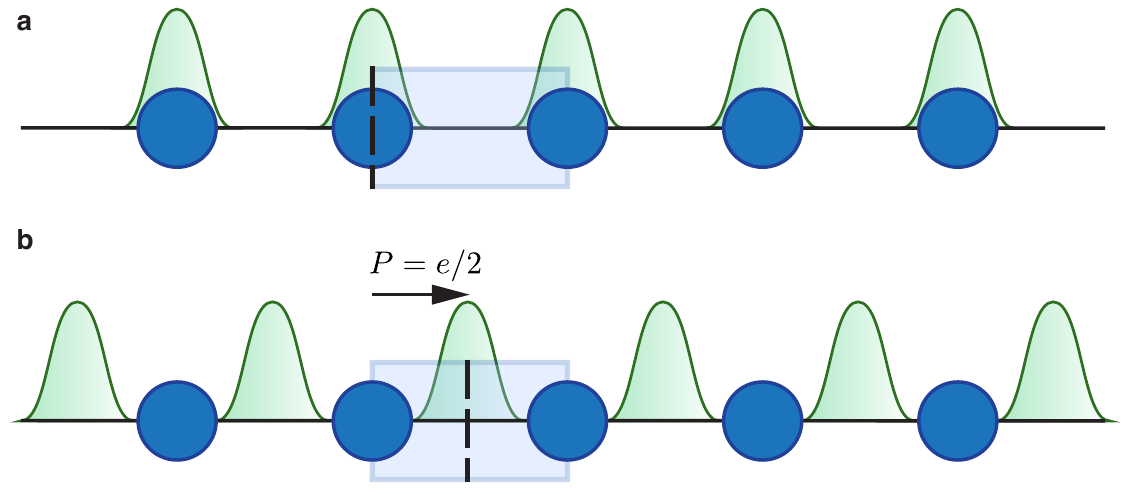}
    \caption{\textbf{Example of one-dimensional topological phases.} \textbf{a} and \textbf{b} show two inversion symmetric positions (dashed lines) of a Wannier center (filled curve) in the unit cell (box). \textbf{a} is an atomic limit, while \textbf{b} is an obstructed topological phase because it cannot be deformed to the atomic limit in \textbf{a} without breaking inversion or closing the band gap. These two configurations differ by a shift of half a lattice site, corresponding to a change in polarization of $P=e/2$ mod $e$.}
    \label{fig:1Dfragile}
\end{figure}
\subsection{\label{sec:basicquant}Basic properties of topological states: atomic obstructions and quantized observables}

A one-dimensional chain with inversion symmetry ($\mathcal{I}$) is an illustrative example of how symmetry can distinguish two topologically distinct states. Consider placing one electron per unit cell in the chain of Fig.~\ref{fig:1Dfragile} with lattice constant $a$. There are two possibilities consistent with $\mathcal{I}$ (recall that $\mathcal{I}: x \to -x$): centering the wave packet at an atomic site (Fig.~\ref{fig:1Dfragile}\textbf{a}), or centering it at the middle point between sites (Fig.~\ref{fig:1Dfragile}\textbf{b}). The first configuration is a good representation of the atomic limit, the wave packet is centred around each atom, and is close to our intuitive picture of a trivial insulator. The second choice is also an insulator, but it is impossible to connect it to the first one respecting $\mathcal{I}$ and simultaneously preserving the gap. Indeed, if we rigidly shift the wave packet away from either of the two inversion-symmetric real space positions in the unit cell (dashed lines) we would preserve the insulating character but break $\mathcal{I}$. If we instead preserve $\mathcal{I}$ by delocalizing the electron symmetrically from one of the two inversion symmetric points we will pass through a delocalized state and close the band gap.

Based on inversion, these two insulators are in different classes; if we respect the inversion symmetry, they are connected by a gap closing. But how can we physically tell them apart? An important property of topological matter is that topologically distinct phases have different physical responses. Since we are discussing inversion symmetry, using the position operator seems like a natural choice. As it turns out, the position operator is central to describe most topological phases, even without inversion. Before we get there, lets consider this operator in our 1D example. 

The expectation value of the position operator over all occupied bands is the polarization, which can be defined, per unit volume, using Bloch's wavefunction $\psi^{n}_\mathbf{k}$ as 
\begin{equation}
\label{eq:Pol}
\mathbf{P}=\dfrac{e}{V}\left\langle \hat{\mathbf{r}}\right\rangle = -\frac{e}{2\pi}\sum_{n\in\mathrm{occ.}}\int_{\mathrm{BZ}} \; d^{D}k \left\langle\psi^{n}_\mathbf{k}\right| i\boldsymbol{\nabla}_\mathbf{k} \left| \psi^{n}_\mathbf{k} \right\rangle.    
\end{equation}
where $V$ is the unit cell volume (equal to the lattice constant $a$ for our 1D system), $\hat{\mathbf{r}}$ is the position operator, and $e$ the electron's electric charge. The integral is to performed over the $D$ dimensional Brillouin zone and summed over occupied bands.
This expression should be intuitively reasonable. The polarization is the average of the wavepacket's position, thus the first equality. For translationally invariant systems we can define the crystal momentum $\mathbf{k}$. Quantum mechanics maps the position operator to a derivative, or a gradient in dimensions higher than one, with respect to the conjugate variable, the momentum $\mathbf{k}$, thus the second equality. Upon a momentum dependent phase shift of the wave function, $\left| \psi^{n}_\mathbf{k} \right\rangle \to e^{i\theta_{\mathbf{k}},n}\left| \psi^{n}_\mathbf{k} \right\rangle$,  you can check, as noted first by Vanderbilt\cite{Resta94}, that the bulk polarization can only change by a unit of electric charge, and thus it is defined only modulo $e$. The two insulators described above are therefore different because their bulk polarizations are $P=e/2$ and $P=0$, mod $e$. This can be proven by direct calculation, but it should also be reasonable because the two insulators differ by a displacement of the electron by half a unit cell (see Fig.~\ref{fig:1Dfragile}). 

Our 1D example already served to illustrate some of the main properties shared by all topological insulators. First, we saw that distinct topological phases can only be connected through a gap closing. Second, topological insulators define quantized observables, which depend only on fundamental constants like $e$, allowing us to classify them. In our 1D example we have distinguished two insulators by their zero or non-zero polarizations, establishing a $0$ or $1$ type of classification, which is mathematically expressed as a $\mathbb{Z}_2$ classification. It is the first example of topological invariant we encounter, because it stays fixed within a topological phase. Third, these observables, and more generally the topological invariants that define them, are determined by global properties of the wavefunctions. Indeed note that Eq.~\eqref{eq:Pol} is an integral over the full Brillouin zone and involves all occupied states. Fourth, these invariants have simple expressions in the presence of symmetry. For example, Eq.~\eqref{eq:Pol} is calculated in momentum space if our system is translationally invariant. More generally, point group symmetries like mirror or inversion can be used to simplify expressions of topological invariants. 

The two one-dimensional insulators in Fig.~\ref{fig:1Dfragile} can be represented in the basis of localized wave functions (our wavepackets were located either at the center of a bond, or at an atomic site). 
The insulators that can be represented in the basis of localized wave functions, but which are not connected to the limit where the wave packet sits at an atomic site are known as bulk obstructed insulators. By this nomenclature Fig.~\ref{fig:1Dfragile}\textbf{a} is an atomic limit,  while Fig.~\ref{fig:1Dfragile}\textbf{b} is a bulk obstructed topological insulator~\eqref{eq:Pol}. A particularly famous system that realizes these phases is the Su-Schrieffer-Heeger chain~\cite{Su79}, a chain of carbon molecules with alternating hopping strength. The ratio between the two hoppings determines whether $P=0$ or $P=e/2$.

\subsection{Strong topological insulators, gapless surface states and symmetry classifications
\label{sec:strongTI}}

All one dimensional insulators are either trivial atomic limits or topological in the sense that they are bulk obstructed insulators, which are not smoothly connected to atomic limits\footnote{We distinguish one-dimensional band insulators from one-dimensional superconductors since the Cooper pairs in topological one-dimensional superconductors are not exponentially localized~\cite{Schindler20}.}. However, one may argue, trivial atomic limits and bulk obstructed insulators are not so different from each other since we can find a basis where the Fourier transforms of Bloch eigenstates, called Wannier functions, are exponentially localized while preserving all the symmetries. This suggests a stricter criterion to label an insulator as topological: a topological insulator is that which cannot be written in a basis of completely localized wavefunctions while preserving all symmetries. Thouless already noticed that it is impossible to ``Wannierize'' a two-dimensional insulator that has a non-zero quantum Hall effect written in terms of Landau levels~\cite{Thouless:1984en}. In 2011, this intuition was extended to three-dimensional insulators~\cite{Soluyanov2011}, and currently this is a criterion used to classify non-interacting topological insulators~\cite{Tang18,Zhang:2019tp,Vergniory:2019ub}.

Topological insulators defined in this way are realized in various forms, depending on the symmetry that protects them\footnote{Two-dimensional insulators with a Hall effect, known as Chern insulators,  are special in the sense that they are protected by the absence of symmetry. They cannot be connected to the atomic limit without a gap closing transition.}. The basic classification of topology relies on 10 symmetry classes\cite{Chiu16} that result from combining the three non-spatial symmetries:  time-reversal, particle-hole and chiral symmetry, the last being the product of the first two. For all these classes, it is possible to define a topological invariant analogous to our polarization example above: it is quantized in integers, determines an observable, and for translationally invariant systems it can be expressed as a global property of Bloch eigenstates in momentum space. Those insulators that realize non-zero integers of these invariants are referred to as \emph{strong} topological insulators.


All surfaces of strong topological insulators are conducting. So long as the symmetry that protects the topological state is respected and the bulk remains open, these conducting surface states cannot be turned insulating, for example by adding impurities. This observation is known as the bulk-boundary correspondence and can be understood as follows.
A boundary of a topological insulator is the collection of points in space between a topologically non-trivial interior (where the bulk topological invariant is non-zero by definition) and the vacuum, which for all intends and purposes is connected to an atomic insulator, which is trivial. If we were to interpolate between these two insulators with a single Hamiltonian, we can choose to vary a parameter that preserves the symmetry. Somewhere along the way, we have to close the gap because these two insulators belong to different topological classes. The place where this occurs is the boundary, and it is gapless. Since this argument only depends on symmetry, and the presence of the bulk gaps, it is insensitive to local details. Thus these gapless states cannot be gapped with local perturbations, such as the presence of impurities, so long as they respect the symmetry that protects the topological state.

Of course in trivial insulators metallic surface states can also exist, but there is nothing fundamental to them. We can always imagine that, by adding local perturbations at the surface, they can be gapped out, and displaced in energy to overlap with bulk energy bands. In contrast, for topological phases, these metallic surface states remain protected against local perturbation, since it is a global, bulk property that guarantees their existence. 

One important example of a three-dimensional strong topological insulator is the $\mathbb{Z}_2$ topological insulator protected by time-reversal symmetry. As before the $\mathbb{Z}_2$ nomenclature refers to the fact that there are two types of insulators in this class, the trivial and the topological, with different electromagnetic responses. In three dimensions, the invariant, in this case $\theta =0 ,\pi$, determines respectively a trivial or quantized magnetoelectric polarizability~\cite{Qi:2008eu,Essin:2009kb}, to which we will return later on in Section \ref{sec:Responsesam}. Bi$_2$Se$_3$ crystals are the paradigmatic example of a $\mathbb{Z}_2$ topological insulator with $\theta = \pi$. 

\begin{figure}
    \centering
    \includegraphics[width=\linewidth]{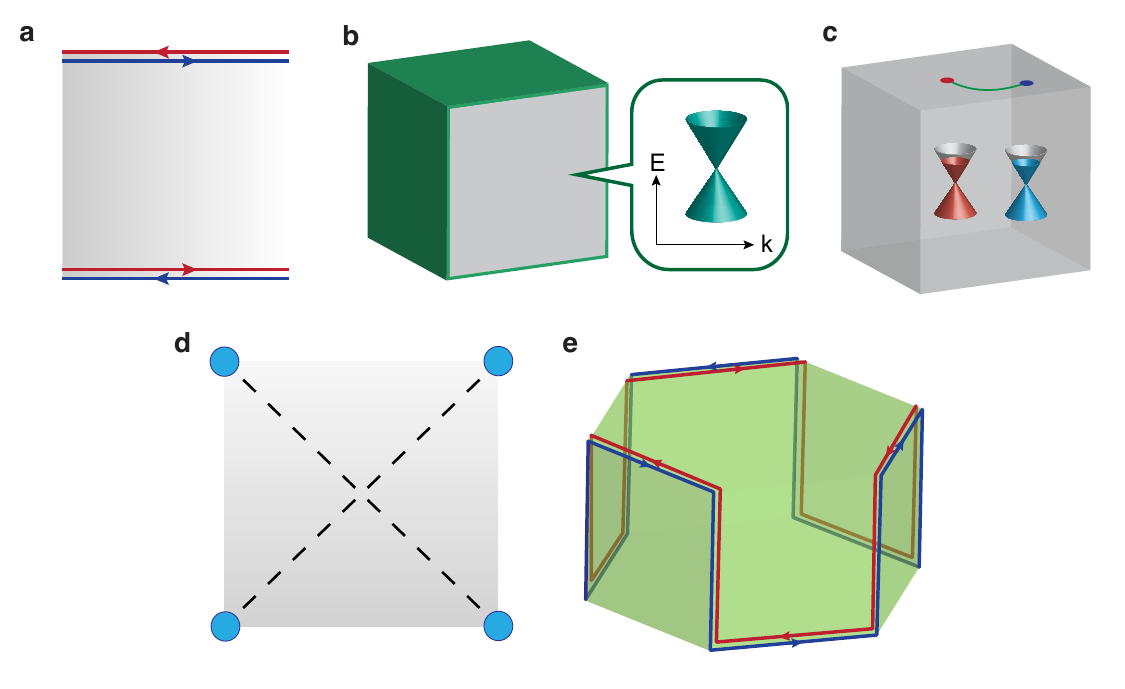}
    \caption{\textbf{Schematics of different topological phases of matter} 
    \textbf{a} $\mathbb{Z}_2$ topological insulator in two dimensions protected by time-reversal symmetry. It features an insulating bulk with two conducting helical edge states, observable for example in nano-SQUID measurements~\cite{Nowack:2013it}. 
    \textbf{b} $\mathbb{Z}_2$ topological insulator in three dimensions protected by time-reversal symmetry. It displays a surface Dirac cone observed in ARPES and spin polarized ARPES showing the spin asymmetry (see for example Ref.~\cite{Jozwiak:2016jv}). \textbf{c} In Weyl semimetals such as TaAs~\cite{Armitage2018}, low-energy quasiparticles have a conical dispersion relation which comes in two flavors, or chiralities. At the surface, the projections of the Weyl nodes are connected by a surface state known as the Fermi arc.
    \textbf{d} An example of a two-dimensional higher-order topological insulators with corner modes (circles) protected by mirror symmetry (dashed lines). They have been observed for example in photonic systems\cite{Mittal:2019tb}. \textbf{e} An example of a three-dimensional higher-order topological insulator with helical hinge states. This state is consistent with what is observed in Bismuth\cite{Schindler:2018hl}.}
    \label{fig:topozoo}
\end{figure}

The gapless surface states of this $\mathbb{Z}_2$ time-reversal invariant topological insulator have remarkable properties. To understand them consider a bulk insulator with time-reversal symmetry. If there is no spin-orbit coupling, time reversal symmetry demands that the bands are even functions of momentum, $E_n(\mathbf{k})=E_n({-\mathbf{k}})$, and that they are doubly degenerate due to spin. This is a consequence of Kramers theorem, that says that two eigenstates that are time-reversal partners of each other must have the same energy. With spin-orbit coupling this degeneracy remains only at time-reversal invariant momenta, where $\mathbf{k} = -\mathbf{k}$ up to a lattice vector (e.g. $\mathbf{k}=(0,0,\pi)$).
Time-reversal symmetry also requires that states at momenta $\mathbf{k}$ and $-\mathbf{k}$ have opposite spin. This is true at the surface too, where the Hamiltonian must be gapless, as we argued above. Combining all of these arguments together the simplest surface Hamiltonian that we can write has two bands and looks like this
\begin{equation}
    \label{eq:Dirac}
    H_\mathrm{surf}(k) = v_\mathrm{F}\mathbf{k}_\parallel\cdot\sigma + \mu \sigma_0.
\end{equation}
Here $\mu$ is a chemical potential, $\mathbf{k}_\parallel = (k_x,k_y)$ is the momentum parallel to the surface and $\sigma_0$ is a $2\times2$ identity matrix.  In the the simplest case the Pauli matrices $\sigma$ represent the spin degree of freedom, but in real materials it may represent a degree of freedom that combines orbital and spin angular momenta. This hamiltonian, known as the Dirac Hamiltonian, is gapless because it describes quasiparticles with a relativistic-like conical dispersion relation $E_\pm(\mathbf{k})=\pm v_{\mathrm{F}}|\mathbf{k}|+\mu$ (see Fig.~\ref{fig:topozoo}\textbf{b}). Since time-reversal symmetry reverts the spin and momentum, a constant $\sigma_z$ term would break this symmetry. The absence of this term keeps the dispersion gapless consistent with our previous bulk based discussion. Additionally, time-reversal symmetry implies that states with opposite momenta have spins pointing in opposite directions.
This requires that the spin rotates, pointing along $\mathbf{k}$, around the Fermi surface, a property known as spin-momentum locking. We will return to these surface properties in Section \ref{sec:expam}.

Another measurable consequence of these arguments is the number of Dirac cones existing at a given surface of a strong $\mathbb{Z}_2$ three-dimensional topological insulator. Note that there can be a Hamiltonian like $H_\mathrm{surf}(\mathbf{k})$ around each time reversal invariant momenta. However, time-reversal symmetry and a non-zero strong topological index requires the number of Dirac cones to be odd~\cite{Hasan:2010kua,Qi2011}. This follows from noticing that if the invariant is non-trivial, the only way to connect the topological insulator bulk and the vacuum, a trivial insulator, is by closing the gap at the surface. It can only close at an odd number of time-reversal invariant momenta, because an even number implies an even number of copies of the surface Hamiltonian Eq.~\eqref{eq:Dirac}, which could be gapped out preserving time-reversal symmetry. Then, this tells us that, without no symmetry other than time-reversal symmetry, $\mathbb{Z}_2$ topological insulators come in two classes, depending if the number of Dirac cones is even or odd ($\theta =0 ,\pi$). An even number of Dirac cones is in this sense is not protected, and disorder respecting time-reversal symmetry can gap them out.

%


It was soon noted that point group symmetries, such as rotations, mirrors or inversion, can play a crucial role in defining and detecting topological phases\cite{FK07,Fu:2011ia}. Firstly, these symmetries can greatly simplify the expressions for strong topological invariants. For example, in the absence of any crystal symmetry, the topological invariant $\theta$ is a complicated integral over the Brillouin zone that involves all filled states. In the presence of inversion symmetry, $\theta$ is given by the product of inversion eigenvalues of all filled bands at all time-reversal invariant momenta~\cite{FK07}, a much more practical expression. This is not an exception.  For instance, the Hall conductivity of two-dimensional insulators with no time reversal symmetry, known as Chern insulators, can be calculated modulo an integer $n$ without any integration if additional $C_n$ rotational symmetries are present. More generally, the use of symmetry to simplify and classify topological phases has lead to the notion of symmetry indicators~\cite{Kruthoff17,Po:2017ci,Po20,Vergniory:2019ub,Zhang:2019tp,Tang18}, which are eigenvalues of point group operations that serve to define and classify different topological phases.


Crystal symmetries can protect topological states, largely enriching the classification of strong topological insulators. These topological insulators are known as crystalline topological insulators~\cite{Fu:2011ia}. In these, only those surfaces that respect the crystal symmetry that protects the phase can hold gapless edge states. A paradigmatic example is SnTe~\cite{Hsieh:2012tq}. It is has a trivial strong topological index but has two Dirac cones at the surface protected by mirror symmetry along the $\left\lbrace110\right\rbrace$ mirror plane. Note that this is an even number of Dirac cones, which as mentioned above are only stable because of the extra mirror symmetry. The effect of disorder on topological crystalline insulators is however a subtle issue and an open research topic.


More generally, there can be boundary states protected in co-dimension higher than one. To understand what this means note that so far we have explained how two (three)-dimensional topological insulators have gapless edges (surfaces). However, point group symmetries can protect gapless corner and hinge states in insulators that go under the name of higher-order topological insulators. For instance, a two-dimensional insulator with two diagonal mirror symmetries can host four corner modes (Fig.~\ref{fig:topozoo}\textbf{d}). In three-dimensions, evidence of hinge-states has been recently found in Bismuth~\cite{Schindler:2018hl} (see Fig.~\ref{fig:topozoo}\textbf{e}).


Although we have not mentioned them explicitly, most of our discussion applies to superconductors described by a Bogoliubov-deGennes Hamiltonian. In this formulation, the Hamiltonian represents a gapped structure with particle-hole symmetry, and as such it is subject to all the above classification schemes. The protected edge states of topological superconductors have a particular significance since particle-hole symmetry requires that there is a doubly degenerate state at zero energy, a combination between electrons and holes. In the particular case of a 1D superconducting chain, there are two doubly degenerate zero modes that are exponentially localized at opposite ends of the chain, known as Majorana zero modes. Two Majorana zero modes may be used to form a q-bit, and since they are topologically protected from decoherence into the bulk so long as the bulk gap remains open, they are a promising platform to realize robust (or topological) quantum computation~\cite{Lutchyn:2018uza}. 
 

The classification of topological phases is still an open and thriving field of research. We note that not all of the properties we have discussed are realized in all topological insulators, yet the fact that distinct topological classes can only be connected by gap closings is universal. Defining all topological phase precisely and generically is currently an open problem. For example recently it was realized that some topological insulators, labeled as \textit{fragile}, can be connected to a trivial state by adding more orbitals to the same lattice structure~\cite{Po:2018,SongZ2019,Song794,Bouhon2019,Peri:2020bn}. The presence of electron-electron interactions, extends the topological classification further into states that are called topologically ordered~\cite{Wen17}. By definition, these states host anyonic excitations, which are quasiparticles with statistics that are neither bosonic nor fermionic. These states can be even further enriched by adding point group symmetries into symmetry enriched topological phases. 

%
\subsection{Topological metals}

The concept of topological stability that we introduced in the previous section seemed to rely on symmetry, but most importantly on the presence of a gap. Are all metals topologically trivial? The answer to this question is a very interesting no. 

Although the first topological gapless state ever discovered was He-3\cite{volovik_book,Bevan:1997vg}, the discovery of topological metals in the solid state\cite{Armitage2018} has fully uncovered their topological nature. 
The simplest class of topological metals are those in which two non-degenerate bands cross close to the Fermi level. These are called Weyl semimetals, and they are topological because if this crossing is isolated from other bands, it is topologically robust. To see this, note that we can write the crossing of two non-degenerate bands in three-dimensions to lowest order in momentum as 
\begin{equation}
\label{eq:Weyl}
H_\mathrm{Weyl}(\mathbf{k}) = v_\mathrm{F}k_x\sigma_x+v_\mathrm{F}k_y\sigma_y + v_\mathrm{F}k_z\sigma_z + \mu \sigma_0,
\end{equation}
which is called the Weyl Hamiltonian. It is gapless at $\mathbf{k}=0$ because the dispersion relation is $E_\mathbf{k}=\pm v_\mathrm{F}|\mathbf{k}|+\mu$. Eq.~\eqref{eq:Weyl} has the very important property that it exhausts all matrices that form the basis in this $2\times2$ subspace of bands, which are the Pauli matrices $\sigma_i$ and the identity $\sigma_0$. Thus, any perturbation restricted to this sub-space will be, to lowest order in momentum, a shift proportional to one of this matrices, say $\mathbf{b}\cdot\boldsymbol{\sigma}+b_0\sigma_0 + \mathcal{O}(k)$. This will only shift the degeneracy point in momentum-energy space, but will not alter the fact that the two bands touch at a point (now at $\mathbf{k}=-\mathbf{b}$). This point is called the Weyl point, or Weyl node (see Fig.~\ref{fig:topozoo}\textbf{c}).

There is a deeper reason for the stability of a Weyl node. The quantity that can reveal this stability is built out of the Berry connection $\mathcal{A}^{n}_{\mathbf{k}} = \left\langle\psi^{n}_\mathbf{k}\right| i\boldsymbol{\nabla}_\mathbf{k} \left| \psi^{n}_\mathbf{k} \right\rangle$. As advertised around Eq.~\eqref{eq:Pol} this quantity is nothing but the position operator coming back again to describe topological states. As before, this quantity is not invariant by a change in the electron's phase $\left|\psi^{n}_\mathbf{k}\right\rangle\to e^{i\theta_{\mathbf{k},n}}\left|\psi^{n}_\mathbf{k}\right\rangle$. This tells us that the Berry connection is not an observable, much like the electromagnetic vector potential is not an observable; we can choose to shift it by the gradient of a scalar leaving Maxwell's equations invariant. Taking further the similarity with the vector potential, we can construct the rotational of $\mathcal{A}^{n}_{\mathbf{k}}$
\begin{equation}
\label{eq:Berrycurv}
    \boldsymbol{\Omega}^{n}_\mathbf{k}=\boldsymbol{\nabla}_\mathbf{k} \times \mathcal{A}^{n}_{\mathbf{k}} = \boldsymbol{\nabla}_\mathbf{k} \times \left\langle\psi^{n}_\mathbf{k}\right| i\boldsymbol{\nabla}_\mathbf{k} \left| \psi^{n}_\mathbf{k} \right\rangle,
\end{equation}
known as the Berry curvature. It is gauge invariant under changes in the phase of the wavefunction, much like a magnetic field is invariant under the choice of different gauges. 
For this reason, the Berry curvature is sometimes referred to as a magnetic field in momentum space.

By direct computation using Eq.~\eqref{eq:Weyl} the Berry curvature around a Weyl node is a monopole in momentum space $\boldsymbol{\Omega}^{n}_\mathbf{k}=\frac{1}{2}\frac{\hat{k}}{|\mathbf{k}|^2}$. Attentive reader, you must frown here! In a lattice system $\left| \psi^{n}_\mathbf{k} \right\rangle$ is periodic, and so should $\boldsymbol{\Omega}^{n}_\mathbf{k}$. The only possible way out is if there is another anti-monopole $\boldsymbol{\Omega}^{n}_\mathbf{k}=-\frac{1}{2}\frac{\hat{k}}{|\mathbf{k}|^2}$ somewhere else in the Brillouin zone. 

That Berry curvature monopoles exist in pairs is a general statement, and a consequence of the Nielsen-Ninomiya theorem: in the band structure of a local Hamiltonian, monopoles of Berry curvature come in pairs of opposite sign\cite{NielNino81a}. This sign is known as the chirality of the node, and so each monopole corresponds to a non-degenerate band crossing with a definite chirality. The chirality of a node is given by the sign of determinant of the velocity matrix sgn(det($M_{ij}$)) where $H = M_{ij} k_i \sigma_j$.  

Due to the restrictions in degeneracy imposed by time-reversal and inversion symmetries, the reader can be easily convinced that at least one of these two symmetries must be broken by the lattice structure to host pairs of Weyl nodes\cite{Armitage2018}. Without time-reversal symmetry the minimum number of nodes is two, while with time-reversal symmetry the minimum number of Weyl nodes is four and inversion is necessarily broken. 

So long as the crossings between opposite chiralities are far in momentum space, they can be treated as independent and remain gapless. The degeneracy can only be lifted if two monopoles of opposite charge are brought together in momentum-energy space. This robustness is topological, since we cannot change the flux of the Berry curvature around a node by adding perturbations in the two-band subspace. This is the reason why the flux over a closed momentum surface that encloses the Weyl point is a topological invariant, called the Chern number. It equals $\pm C$ where $\pm$ is determined by the chirality and $C$ is the monopole charge or Chern number. In our example Eq.~\eqref{eq:Weyl}, $C=1$.

We arrive then to the definition of a Weyl semimetal, which is that material where the two nodes are close to the Fermi energy, but separated in momentum space. But why are these crossings important? The existence of these node is fundamentally interesting because electrons close to it behave as pseudo-relativistic particles with no mass, like a photon, but with electrical charge. In this sense the quasiparticles in Weyl semimetals behave similarly to those in graphene but moving in three-dimensions, and topologically protected. One of the main differences with graphene, that we shall not prove, is that the existence of Weyl nodes implies the existence of robust chiral surface states, known as Fermi arcs, that cannot be removed. In an heuristic way, these can be thought of as the Dirac string that connects both monopoles, manifesting itself at the surface of the material (see Fig.~\ref{fig:topozoo}\textbf{c}).

When coupled to an electromagnetic field the Weyl Hamiltonian Eq.~\eqref{eq:Weyl} has a number of important physical consequences, some of great technological potential. A particularly surprising one is that under applied parallel electric and magnetic fields, the balance between the number of quasiparticles of different chiralities is destroyed. This phenomena is known as the chiral anomaly in high-energy physics and in condensed matter it significantly enhances the magnetoconductivity\cite{Armitage2018}. Weyl quasiparticles have significantly different responses to light, including a universally quantized photocurrent\cite{deJuan:tma}. 

Weyl semimetals were discovered in the solid state in 2014 first in TaAs, and by now many other materials have been identified or predicted to be in this phase~\cite{Armitage2018}. Weyl semimetals exist both in magnetic and non-magnetic materials, and Fermi arcs, enhanced longitudinal magnetoresistence and other signatures are by now routinely observed in experiment.

Finally, as with topological insulators, point group symmetries can enhance the protection, and enrich the classification of topological semimetals. For example, in Dirac semimetals like Na$_3$Bi two chiralities meet in momentum space without opening up a gap due to additional lattice symmetry that prevents it\cite{Armitage2018}. These symmetries can also protect nodal line crossings, for example in non-symmorphic materials with negligible spin-orbit coupling. Recently, crossings of more than two bands, known as multifold fermions\cite{Manes:2012fi,Chang:2018uia,Bradlyn2016}, have been shown to be stabilized by screw rotations and other non-symmorphic symmetries. These have been found to exist in RhSi\cite{Chang2017,Tang2017}, CoSi\cite{Tang2017}, and AlPt\cite{Schroter:2019wqa}, among other materials, and share many of the exciting and useful electromagnetic responses of Weyl semimetals~\cite{Flicker2018Chiral}. 

\subsection{Topological phases in non-electronic systems}
 
Topological properties transcend electronic systems into the realm of the synthetic, the macroscopic and the classical. As we have discussed, topological properties constrain the existence of degeneracies in band structures. Therefore, topological states can occur in any system where we can define a band structure: sound modes in acoustic 3D printed metamaterials\cite{Peri:2019hn}, waveguide modes in photonic metamaterials~\cite{Ozawa:2019ij} (see Fig.~\ref{fig:topozoo}\textbf{c}), ultra-cold atomic lattices~\cite{Cooper:2019hv}, or the energy-phase relation in multi-terminal Josephson junctions\cite{Riwar:2016hr}.

Topological states can be defined even for classical systems. Recently, the mechanical theory of stress and strain in mechanically constrained objects has also been connected to topological states~\cite{Kane:2014if}. For example, a mechanically constrained one-dimensional chain of rods can realize a mechanical analogue of the Su-Schrieffer-Heeger model~\cite{Su79} mentioned above. Perhaps even more striking is the connection of equatorial waves on Earth to topological boundary modes allowed by the Earth's rotation, and the consequent breaking of time-reversal symmetry\cite{Delplace:2017kq}. 

Topological states seem ubiquitous. Our understanding of them is strongly centred on lattice periodicity. The remainder of this chapter dives into the fate of topology in the absence of translational invariance, and the striking recent progress made to understand it.

\section{Topological phases beyond crystalline solids}

In our previous discussion we purposely highlighted how periodicity, and symmetry in general, simplify the discussion, classification and computation of topological properties. Expressions of topological invariants are succinctly written in momentum space. Crystalline symmetries simplify further these expressions and define new topological states. Is this all lost when the symmetry under lattice translations is broken? Although the reader might have guessed that the answer is no, one could still ask, is it worth the effort to find out? If topological matter exists already, why bother with translationally non-invariant systems? Here we give two reasons to keep on reading.

The first reason is conceptual. Topological states are robust, but how robust? We will see that on-site disorder on a crystalline lattice can induce topological phases, but eventually will lead to a topologically trivial Anderson insulator. Can the absence of a crystalline lattice support topological states as well? If so, how can we find them systematically? and, are they different from topological states of matter found in crystals? Even more fundamentally, how can we define and calculate invariants that signal the presence of non-crystalline topological states in general, and with them their physical properties? More viscerally, what on earth does it mean to say we have a Dirac cone on the surface of a strong amorphous topological insulator if there is no momentum-space left to define it?

The second reason is technological. Topological insulators hold a great technological promise, rooted in their special properties, like their robust spin-momentum locking at the surface, useful for spin-tronic devices, or topological superconductivity, which can open the doors to quantum computation. However, it is experimentally challenging to grow topological insulators where the chemical potential is in the gap, clearly exposing the topological properties. For example, heroic efforts were needed to turn the paradigmatic Bi$_2$Se$_3$ into a bulk insulator~\cite{Wu1124}. Moreover, it is desirable that topological properties are scalable. Among many things, this could mean that the growth process of topological solids should be as simple as possible, while preserving the technologically useful properties. If all topological properties are observed only in carefully grown crystals, production costs might rise and technological impact would be less likely. Amorphous solids are easy to grow, one of the reasons behind their ubiquity in technology. Since translational symmetry is not a requirement for topological protection, combining amorphousness with useful topological properties seems a likely recipe for new states, new phenomena and new applications. 

The statement that amorphous matter can be topological may not be so surprising after all. We know that another key classification scheme, symmetry breaking, does not depend on crystallinity. For example, being a solid has nothing to do with being a crystal. A solid can be defined by its response properties to elastic shear, and since amorphous phases can present such properties, those that do can be called solids. Similarly, topological phases are characterized by their responses to external perturbations, encoded mathematically in the form of topological invariants. Although both solidity and topology are conveniently expressed exploiting the benefits of periodicity, they do not rely on crystallinity to exist, despite their shared absence in classical solid-state textbooks \cite{Ashcroft76}.

\begin{figure}
    \centering
    \includegraphics[width=\linewidth]{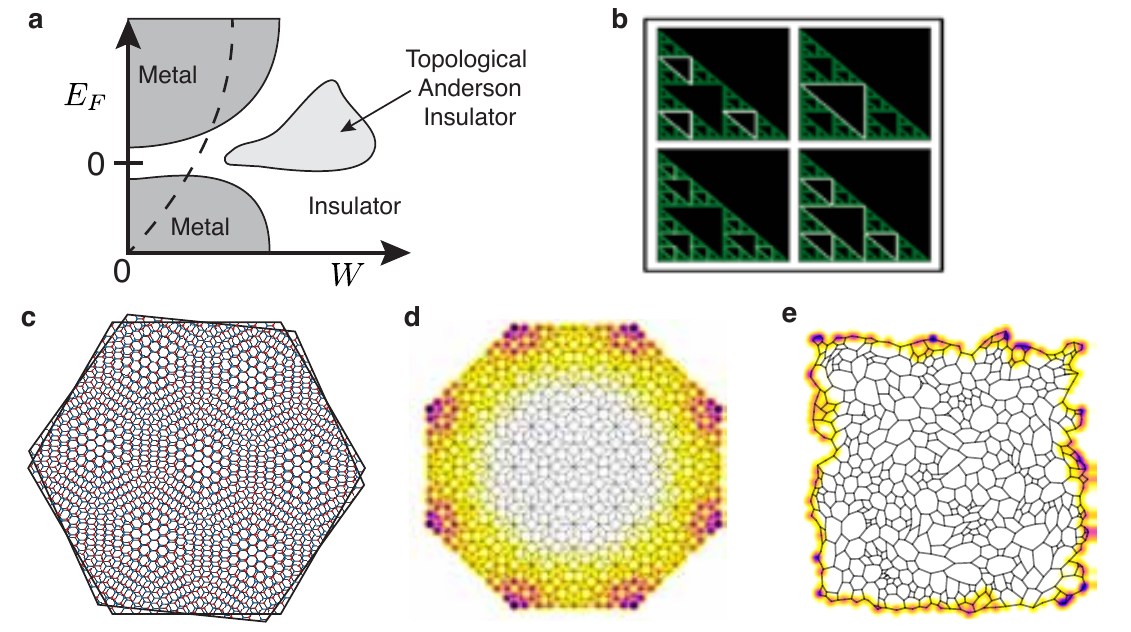}
    \caption{\textbf{Examples of topological phases of matter in systems without translational symmetry} \textbf{a} Schematic phase diagram of a topological Anderson insulator. As disorder $U_0$ is increased the average conductance plateaus at $G=2e^2/h$, signaling a two-dimensional topological phase with two conducting edge channels known as topological Anderson insulator\cite{Groth09}. The dashed line signals where the effective topological gap inverts from trivial to topological. 
    \textbf{b} Topological edge states in a fractal Sierpinski gasket under a magnetic field (adapted with permission from Ref.\cite{Brzezi18}).
    \textbf{c} Two layers of graphene twisted by a small angle $\theta$ can host a variety of topological phases including Chern insulators and higher-order topological phases with corner modes. 
    \textbf{d} Higher-order quasi-crystalline topological insulator with corner states protected  by $C_8$ symmetry (adapted with permission from Ref.\cite{Varjas2019}). \textbf{e} Amorphous topological Chern insulator phase showing the gapless edge state (adapted with permissison from Ref.\cite{Marsal20}).}
    \label{fig:aperiodictopozoo}
\end{figure}

\subsection{Half a step towards amorphous topological matter: disorder, quasicrystals and moir\'{e} lattices}

The question of the robustness of topological states under disorder is as old as the field of topological phases itself. It is now understood that topological properties do not require a crystalline lattice, and can survive or even be enhanced by disorder. 

A clear example is that of the topological Anderson insulator~\cite{Jiang09,Groth09,Jian09}. To understand it, take a step back and recall some of the main properties of the Anderson insulator state, which arises in a disordered solid. Anderson modelled a disordered solid by considering a randomly varying onsite potential drawn from a distribution of finite width $w$, representing the disorder strength. Upon increasing $w$ solids become insulators with exponentially localized wavefunctions. The susceptibility to localize depends, among other things, on the dimensionality of the systems in question: an infinitesimal amount of disorder localizes 1D states while in 3D systems a strong disorder strength is needed.

Are topological states different? Although strong disorder is expected to close the bulk gap and destroy the topological phase, topological properties can remain robust against a significant amount of disorder. For example, if disorder does not break the symmetry that protects the state, adding weak to intermediate disorder will not destroy the topological state. The most striking and technologically relevant example is once more the strong $\mathbb{Z}_2$ insulator protected by time reversal invariance. If disorder is non-magnetic, it preserves time-reversal symmetry, and back-scattering is forbidden. 

There is an even more surprising case, in which disorder acts, for once, as an ally. The topological Anderson insulator is a state in which disorder, in moderate amounts, has turned a trivial insulator into a topological one. It happens when the trivial insulator is on the verge of turning into a topological insulator. On-site disorder might tip the balance by renormalizing the onsite energy such that the trivial insulator is pushed, through a gapless state into a topological insulator (see Fig.~\ref{fig:aperiodictopozoo}\textbf{a}). If disorder is strengthened even further, the topological Anderson insulator falls into a trivial Anderson insulator. This phenomenon has been observed in disordered atomic wires of trapped ultra-cold atoms~\cite{Meier929}. 

The search for topological states without crystalline order goes beyond the study of random disorder.
Quasicrystals are a recent addition to the collection of systems that can display topological phases~\cite{Kraus:2012iqa,Tran15,Fuchs2016,Bandres:2016gx,Fulga:2016jo,Fuchs2018,Huang:2018bx,Huang:2018gu,Varjas2019,Chen2020,Chen19}. They are ordered aperiodic structures, that show sharp diffraction peaks which are incompatible with lattice translations~\cite{Shechtman:1984kf}. Analogous to two-dimensional crystals, two-dimensional quasicrystals are predicted to display a quantum-Hall effect~\cite{Tran15,Fuchs2016,Fuchs2018}, which can be engineered in photonic lattices by periodically modulating waveguides~\cite{Bandres:2016gx}, two-dimensional topological superconductivity~\cite{Fulga:2016jo}, two-dimensional quantum spin Hall phase~\cite{Huang:2018bx,Huang:2018gu}, and a topological Anderson insulator phase~\cite{Chen19}. Notably, there are new topological phases predicted to occur only in quasicrystals~\cite{Varjas2019,Chen2020}. For example, eightfold rotations, possible only in quasicrystalline lattices, can stabilize Majorana corner modes~\cite{Varjas2019} in a higher-order topological quasicrystal (see Fig.~\ref{fig:aperiodictopozoo}\textbf{d}). Additionally, some type of quasicrystals can be constructed from projecting sites on a higher-dimensional crystal to a line or plane, sometimes referred to as the cut-and-project method~\cite{Tran15}. This relationship to regular crystals maps certain quasicrystalline topological phases into crystal ones, yet allowing in principle to reach higher-dimensional invariants beyond three dimensions. This ideas have been explicitly demonstrated in a one-dimensional Harper lattice realized in a photonic system~\cite{Kraus:2012iqa}.

Topological states of matter are rapidly expanding to other systems without strict translational invariance. For example, it is currently investigated if topological phases can exist in fractal lattices (Fig~\ref{fig:aperiodictopozoo}\textbf{b}), which have non-integer dimensionality. 
Numerical evidence supporting this possibility is currently being pursued in the particular case of quantum Hall like states~\cite{Agarwala:2018vc,Brzezi18,Pai19,Fremling20}. Although promising, the numerical simulations of fractals are necessarily of finite depth. A rigorous proof that fractals can host topological states is still absent. 

Finally, we comment on another related and exciting search frontier: twisted bilayer graphene~\cite{Cao:2018tp,Cao:2018wy}. This system is composed of two perfectly hexagonal lattices, twisted by a small angle $\theta \sim 1.1^o$ with respect to each other. This forms a moir\'{e} superlattice that although not strictly periodic, it is close enough to treat the moir\'{e} supercell as periodic (see Fig.~\ref{fig:aperiodictopozoo}\textbf{c}). At low energies the large real space unit cell folds the original Brillouin zone where the original bilayer graphene bands overlap. A minimal tight-binding model\cite{Bistritzer12233} predicts that the band width of the relevant bands close to half-filling, graphene's natural state, are extremely flat, reaching close to exact flatness at certain angles known as magic angles (including $\theta \sim 1.1^o$). The band flatness enhances the effect of interactions compared to the typical kinetic energy scale, set by the small band-width. Twisted bilayer graphene displays a set of interaction induced phases, such as superconductivity~\cite{Cao:2018tp,Cao:2018wy}, but also a quantized Hall effect in the absence of magnetic fields\cite{Serlineaay5533}, known as the quantum anomalous Hall effect.

The above serves to emphasize how, without the constraints of translational invariance, topological phases can display a richer set of phenomena. Given the above it is only natural to expect that amorphous solids can also display topological phases (see Fig.~\ref{fig:aperiodictopozoo}\textbf{e}), with the additional benefit of being ubiquitous.

\subsection{Modeling amorphous topological matter \label{sec:modelsamorp}}

\begin{figure}
    \centering
    \includegraphics[width=\linewidth]{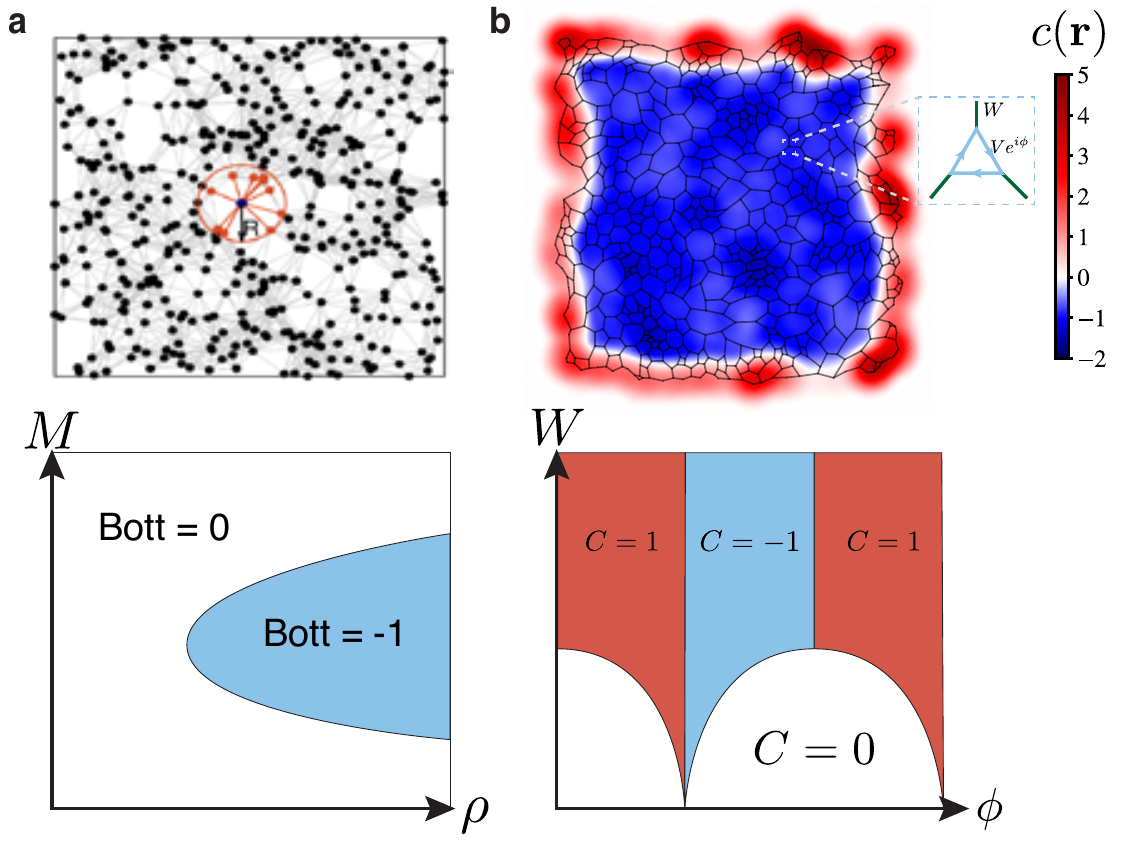}
    \caption{\textbf{Examples of amorphous models of topological matter} \textbf{a} Random point set connected by hopping up to a radius $R$. The Chern insulator phase is diagnosed by localized edge states and the Bott index, Eq.~\eqref{eq:Bott}, shown in the lower panel (adapted from  Ref.~\cite{Agarwala:2017jv}). \textbf{b} Topological Weaire-Thorpe model with threefold coordination. The topological phase is diagnosed by a quantized local Chern marker, Eq.~\eqref{eq:LCM}, in the bulk of the sample. The edge states contribute with a large and positive Chern marker such that the average over the sample is zero. The topological phase diagram (bottom) can be inferred analytically using amorphous symmetry indicators (adapted from Ref.~\cite{Marsal20}).}
    \label{fig:amorphous models}
\end{figure}

%
%
Amorphous topological states were proposed independently by two groups at the dawn of 2017~\cite{Mitchell2018,Agarwala:2017jv}. As a proof of principle, Agarwala and Shenoy\cite{Agarwala:2017jv} modeled an amorphous solid with a tight-binding hopping model using an uncorrelated uniform distribution of sites. They chose the Hamiltonian in the form
\begin{equation}
    H = \sum_{i\alpha}\sum_{j\beta} (t(r)T_{\alpha\beta}(\hat{\mathbf{r}})+\varepsilon_{\alpha\beta}\delta_{ij}) c^{\dagger}_{i\alpha}c_{j\beta},
\end{equation}
that specifies the hopping $t(r)T_{\alpha\beta}(\hat{\mathbf{r}})$ between $ij$ sites and $\alpha\beta$ orbitals, with an on-site potential term $\varepsilon_{\alpha\beta}$. They imposed that the hopping amplitude is finite in a certain characteristic radius by choosing $t(r)= C\Theta(R-r)e^{-r/a}$ (see Fig.\ref{fig:amorphous models}\textbf{a}). The topological phase diagram is dictated by the angular part of the hopping, $T_{\alpha\beta}$. To define it they were mainly inspired by two types of topological insulators models defined for crystal lattices: the three-dimensional $\mathbb{Z}_2$ time-reversal invariant topological insulator described above, and the Chern insulator, a two dimensional system with a quantized Hall conductivity. As a pedagogical exercise we now describe the latter example in some detail.

The term Chern insulator refers to any two-dimensional insulator that has a finite Hall conductivity. This last property necessarily means that time-reversal symmetry is broken, either by an external magnetic field, or intrinsic magnetic moments. The Chern insulator is a special kind of topological insulating state because although it cannot be deformed to an atomic insulator without closing the gap (as our definition of topological state requires) there is no symmetry that protects it. Semiclassically, the absence of time-reversal symmetry allows cyclotron orbits to form, as in the Hall effect. Although seemingly localized in the bulk, at the edges they result in skipping orbits that cannot back scatter. Quantum mechanically this translates into an obstruction to find a (Wannierized) basis of localized orbitals if the Hall conductivity is finite, as discussed in Section~\ref{sec:strongTI}. 

Since the Chern insulator does not require any symmetry, it is a natural starting point to construct a topological insulator. But how do we build such a model? 
On a periodic lattice the Hall conductivity of a Chern insulator is given by the integral of the Berry curvature, Eq.~\eqref{eq:Berrycurv}, over the Brillouin zone, summed over all filled bands~\cite{Haldane04} 
\begin{equation}
\label{eq:sigmaxy}
    \sigma_{xy} = \dfrac{e^2}{h}\sum_{n\in\mathrm{filled}}\int \dfrac{d^2k}{(2\pi)^2}\Omega^{n}_{z}(\mathbf{k}) = Ce^2/h.
\end{equation}
Note that since in two dimensions we only have two momenta, $k_x$ and $k_y$, the Berry curvature only has one component, along $z$. The integer $C$ is the Chern number, already encountered above. For the purposes of proving that it is indeed an integer, we shall restrict to two-band models, yet the statement is independent of the number of bands. Any two band Hamiltonian defined in a periodic lattice has the general momentum space form $H(\mathbf{k})=\mathbf{d}_{\mathbf{k}}\cdot\boldsymbol{\sigma}+\varepsilon_\mathbf{k}\sigma_0$ because the three Pauli matrices and the identity form a complete basis of $2\times 2$ hermitian matrices. This implies that $\mathbf{d}_{\mathbf{k}}$ is a real vector function of momentum, $\varepsilon_\mathbf{k}$ is a different function of momentum, and they are all periodic. Given this form, it is well understood how to choose $\mathbf{d}_{\mathbf{k}}$ and $\varepsilon_\mathbf{k}$ to build a topological insulator with finite Hall conductivity. The intuition is as follows. For such a two-band model Eq.~\eqref{eq:sigmaxy} can be rewritten as
\begin{equation}
\label{eq:sigmad}
    \sigma_{xy} = \dfrac{e^2}{h}\int \dfrac{d^2k}{(2\pi)^2} \hat{d}_{\mathbf{k}} \cdot (\partial_{k_x}\hat{d}_{\mathbf{k}}\times \partial_{k_y}\hat{d}_{\mathbf{k}}),
\end{equation}
where we defined the unit vector $\hat{d}_{\mathbf{k}}=\mathbf{d}_\mathbf{k}/|{\mathbf{d}_{\mathbf{k}}}|$. The mathematically inclined reader, or those familiar with magnetic textures can recall that the integral is quantized to integers over a closed manifold, in this case the Brillouin zone. This number, sometimes referred to as the skyrmion number, counts how many times the unit vector $\hat{{d}_{\mathbf{k}}}$ covers the surface of the sphere that it defines as we vary $\mathbf{k}$. Thus, it suffices to see how many times the vector $\hat{{d}_{\mathbf{k}}}$ reaches the north and south poles ($\hat{d}_{\mathbf{k}}=(0,0,\pm1)$). 

We can now use the above arguments (or direct computation) to show that a simple model that interpolates between a trivial insulator and one with a finite Hall conductivity is
\begin{equation}
\label{eq:CImodelk}
    H = t_1\sin(k_x)\sigma_x+t_1\sin(k_y)\sigma_y+(M+2t_1-t_1\cos(k_x)-t\cos(k_y))\sigma_z.
\end{equation}
If $M$ is large and positive, $\hat{{d}_{\mathbf{k}}}$ remains close to $(0,0,1)$ for all $\mathbf{k}$. Since $\hat{{d}_{\mathbf{k}}}$ does not cover the sphere it defines a trivial insulator with $\sigma_{xy}=0$, and thus $C=0$. If $0<(M+2t)<2$ however, we notice that $\hat{d}_{\mathbf{k}=(0,0)}=(0,0,-1)$. As we move the momentum the orientation of $\hat{d}_{\mathbf{k}}$ changes until it reaches $\hat{d}_{\mathbf{k}=(0,\pi)}=\hat{d}_{\mathbf{k}=(\pi,0)}=\hat{d}_{\mathbf{k}=(\pi,\pi)}=(0,0,1)$. Therefore $\hat{d}_{\mathbf{k}}$ covers the sphere exactly once as we move $\mathbf{k}$ in the Brillouin zone, defining a Chern insulator with $\sigma_{xy}=e^2/h$, and thus $C=1$.

In real space Eq.~\eqref{eq:CImodelk} can be defined on a square lattice with two orbitals per site
\begin{eqnarray}
    \nonumber
    H &=& \frac{t_1}{2}\sum_{i}\left[c^{\dagger}_{i}\left(i\sigma_x-\sigma_z \right)c_{i+\hat{x}}+c^{\dagger}_{i}\left(i\sigma_y -\sigma_z \right)c_{i+\hat{y}}+\mathrm{h.c.}\right]\\
    &+& (2t_1+M)c^{\dagger}_{i}\sigma_z c_{i},
    \label{eq:CImodelx}
\end{eqnarray}
where the creation and annihilation operators are two component vectors in orbital space (e.g. $c^{\dagger}_{i}=(c^{\dagger}_{i,\alpha},c^{\dagger}_{i,\beta})$), and $\hat{x},\hat{y}$ are unit vectors. 

Agarwala and Shenoy generalized these hoppings from a square lattice to a random point set. To do so it is illustrative to write the onsite matrix $\varepsilon_{\alpha \beta}$, and hopping matrix in the $x$ direction $T^{(x)}_{\alpha \beta}(\theta)$ connecting two sites $i$ and $j$
\begin{equation}
    \varepsilon_{\alpha \beta}=\left(
    \begin{array}{cc} 2t_1+M & 0 \\
    0 & -(2t_1+M)
    \end{array}\right), \hspace{2mm}
    T^{(x)}_{\alpha \beta}=\frac{t_1}{2}\left(
    \begin{array}{cc} -1 & -i \\
    -i & 1
    \end{array}\right)\equiv T_{\alpha \beta}(\theta=0).
\end{equation}
The latter definition, where we introduced the polar angle $\theta$, suggests how to generalize this hamiltonian to arbitrary $\theta$, provided we identify the hoppings in $x$ direction with $\theta=0,\pi$ and the hoppings in $y$ direction with $\theta=\pm\pi/2$. Hermiticity of the Hamiltonian demands that $T_{\alpha \beta}(\theta)=T^*_{ \beta\alpha}(\theta+\pi)$. This results in
\begin{equation}
    T_{\alpha \beta}(\hat{\boldsymbol{r}})= \frac{1}{2}
    \left(\begin{array}{cc}
    -t_1+t_{2} & -i t_1e^{-i \theta}+f(\theta) \\ 
    -i t_1 e^{i \theta}+f^*(\theta) & t_1+t_{2}
    \end{array}\right).
\end{equation}
The phase factor $e^{-i \theta_{ij}}=(x_{ij}+iy_{ij})/|\mathbf{r}_{ij}|$, with $x_{ij}=x_{i}-x_{j}$ (resp. for $y$ and $\mathbf{r}$), captures the angle dependence of the vector connecting two sites $ij$. The Hamiltonian in Eq.~\eqref{eq:CImodelx} is recovered when we exclude $f(\theta)$ and $t_2=0$, not present in Eq.~\eqref{eq:CImodelx}, and only allow for vertical and horizontal hoppings. To break particle-hole symmetry, Argawala and Shenoy~\cite{Agarwala:2017jv} allowed the constant $t_2$ to be finite, and to take into account inter-orbital mixing they incorporated an even function of the angle $f(\theta)=\lambda\left[\sin ^{2} \theta(1+i)-1\right]$. The latter effect also modifies the onsite energy as
\begin{equation}
    \boldsymbol{\epsilon}_{\alpha \beta}=
    \left(\begin{array}{cc}
    2t_1+M & (1-i) \lambda \\ 
    (1+i)\lambda & -(2t_1+M)
    \end{array}\right),
\end{equation}
which completes the definition of the Agarwala and Shenoy model\cite{Agarwala:2017jv}. 

With periodic boundary conditions this model shows gaps as a function of $M$, as expected from our crystalline discussion. For open boundary conditions and intermediate values of $M$ the gaps become filled with states that are localized at the one-dimensional boundaries of the system, suggesting their topological origin. By computing topological markers, which are observables that signal topological phases for disordered systems and are described in the next section, they showed that these models could indeed host topological phases for high enough atomic densities, defined as $\rho = N_{\mathrm{sites}}/V$, where $N_{\mathrm{sites}}$ is the number of sites and $V$ is the volume (see Fig.~\ref{fig:amorphous models}\textbf{a}).

The above recipe, is generalizable to representative models for all of the 10 strong topological symmetry classes~\cite{Agarwala:2017jv}. Indeed bulk gaps exist where metallic surface states of putative topological origin survive. We use putative because, as we will mention later when we discuss how to detect these phases, it is not known how topological markers generalize to all symmetries. 

Concurrently with the work by Argawala and Shenoy, Mitchell \textit{et al.}\cite{Mitchell2018} theoretically proposed and experimentally demonstrated the possibility of topological insulating phases in two dimensions using coupled gyroscopes. They demonstrated experimentally that several types of amorphous networks, built out of different point sets, including hyperuniform, jammed, quasi-crystalline and uniformly distributed, can support edge states that propagate chirally along the edges of the sample. 

Beyond the proof of principle, and remarkable experimental realization\footnote{We encourage the reader to view the supplementary videos in Ref.~\cite{Mitchell2018} for a fascinating visualization of the edge state dynamics.}, their findings showed that many, but not all networks, hosted topological phases. To define amorphous lattices Mitchell \textit{et al.} defined first a point set. The point sets were of four types: randomly distributed sites (as in the Argawala and Shenoy model), hyperuniform, jammed, and  quasicrystalline. Hyperuniform point~\cite{Florescu:2009ev} sets are defined as those sets where the number of points in a given observation window decays faster than the window's area.  This is quantified by the structure factor $S(\mathbf{k})=\frac{1}{N}\sum^{N}_{i}|\mathrm{exp}(i\mathbf{r}_i\cdot \mathbf{k})|^2$ for $k\neq 0$ demanding that it grows as $S(\mathbf{k})\propto k^{\alpha}$ as $k\to 0$ with $\alpha>0$. Jammed point sets are defined by minimizing a potential interaction (in this case a harmonic potential) between non-uniformly distributed (polydisperse) discs. For a given point set, Ref.~\cite{Mitchell2018} applied different connectivity rules to define different amorphous lattices.  By connecting the nearest-neighbours of a given point they arrived to an amorphous lattice formed by triangles of connected sites, known as Delaunay triangulation. Connecting the centers of the Delaunay triangles they defined the Voronized, threefold-coordinated lattice. Decorating each site in a Voronized lattice with a triangle we arrive to a Kagome decoration of the Voronized lattice. 

All but Delaunay triangulations showed gapped spectra and topological phases~\cite{Mitchell2018}. This observation was interpreted as the amorphous counterpart to the fact that gapped, topological phases need more than two bands to be realized. The Delaunay triangulation locally resembles a triangular lattice which, all sites being equal, is gapless. In contrast, the Voronoi tessellation locally resembles the honeycomb lattice, which can host topological phases. As noted early on in the amorphous community~\cite{Weaire71} local geometry of the lattice plays a crucial role in determining band gaps, and consequently realizing topological phases. We will return to this point shortly.

The works by Agarwala and Shenoy\cite{Agarwala:2017jv}, and Mitchell et al.\cite{Mitchell2018} opened up the door to topological phases in amorphous lattices in various physical systems. Soon after, amorphous photonic lattices with variable packing fractions formed by jammed point sets~\cite{Mansha:2017be} and uniform point sets~\cite{Xiao:2017ji} were proposed to host chiral edge states. Amorphous two-dimensional topological superconductivity was proposed to emerge in a Shiba glass, arising from hybridizing states formed around magnetic moment on a superconducting surface with strong Rashba spin-orbit
coupling~\cite{Poyhonen2017}. Using tight-binding molecular dynamics it was also proposed that liquid states can inherit topological properties upon melting a regular lattice~\cite{Chern:2018gvb}. Finally, starting from model tight-binding Hamiltonians, the approach of Agarwala and Shenoy described above serves to generalize amorphous topological phases to amorphous topological metals~\cite{Yang19} and higher-order topological phases~\cite{agarwala2019higher}.


Despite the above breakthroughs, many of the tight-binding models based on random point sets or exponentially decaying hopping matrices ignore that amorphous solids often do have a local environment very similar to that of a crystal, with a fixed coordination number. In other words, amorphous matter is not entirely random. In covalently bonded amorphous solids the local correlation is similar to that of the crystal. The coordination number remains fixed, nearest-neighbour distances are strongly peaked around the crystal bond lengths, and there are no dangling bonds\cite{Zallen}. Amorphous solids differ from the crystal in the absence of long-range correlation and the significant spread in bond angles, but the local environment remains fairly similar to the crystal.

The coordination number in particular conveys a significant amount of chemical and topological information, determining the local environment of a site. In the 1970s this observation motivated Weaire and Thorpe to construct a model with fixed tetrahedral coordination to describe amorphous silicon~\cite{Weaire71}. For generic coordination $z$ it reads
\begin{equation}
    H = \sum^{z}_{i,j\neq j'} V^{(i)}_{jj'}\ket{i,j}\bra{i,j'} + \sum^{z}_{i\neq i',j} W^{(j)}_{ii'} \ket{i,j}\bra{i',j}.
    \label{eq:Topo_WT}
\end{equation}
The index $i$ labels the sites within a $z$ coordinated lattice, and the $j$ index labels the $z$ orbitals within a site  ($j=1,2,...z$), see Fig. \ref{fig:amorphous models}\textbf{b}. Weaire and Thorpe chose the matrices $V^{(i)}_{jj'}= V$ and $W^{(j)}_{ii'}=W$ to be real, and independent of the site $i$ and orbital $j$ respectively.
Using the resolvent method~\cite{Schwartz72} it is possible to show that for any generic $V$, $W$ and $z$ this model has band gaps where energy states are forbidden to appear. If the filling is such that it falls in one of these band gaps, the model describes an insulator. Although these models do not capture the fine spectral details of the density of states they capture the general feature that covalently bonded amorphous are generically gapped. 

Due to their gapped structure it is then natural to consider the Weaire-Thorpe Hamiltonians as descriptions of topological insulating phases in amorphous solids. This route was explored recently~\cite{Marsal20} by allowing the intra-site coupling $V$ to be complex by replacing it with $Ve^{i\phi}$ (see inset of Fig.~\ref{fig:amorphous models}\textbf{b}). Using the resolvent method it is possible to prove that the model remains gapped. For a two-dimensional version without time-reversal symmetry the model hosts different Hall insulating phases as a function of the magnitude and phase of the intra-site hopping ($V$ and $\phi$), and the inter-site hopping ($W$), proving that this model can indeed be topological for commensurate fillings. The authors noticed that the equivalence between orbitals and the fixed coordination of this model can be exploited further to reproduce the phase diagram without the need to calculate a topological invariant. We will return to this aspect in the next section, when we discuss how to detect topological phases.

%
%
In addition to tight-binding models, more sophisticated numerical approaches are powerful tools to generate amorphous topological lattices. Density functional theory predicts\cite{Costa:2019kc} that the $\mathbb{Z}_2$ topological insulator character of crystal bismuthene~\cite{Reis287}, a monolayer of bismuth recently synthesized, retains its topological character in its amorphous form.
Molecular dynamics simulations have also been employed to model amorphous topological insulators. In its crystalline form Sb$_2$Te$_3$ is a topological insulator~\cite{Zhang:2009ks,Hsieh09,Pauly12} and using molecular dynamics its amorphous counterpart is predicted to be gapped~\cite{Caravati10}. However the survival of topological properties was not discussed in this work. Supported by the experiments described below, molecular dynamics simulations indicate that a gap of $293$ meV exists in amorphous Bi$_2$Se$_3$, similar in magnitude to that of the crystal ($\sim 300$ meV), and consistent with the thermally activated resistivity curves ~\cite{Corbae:2019tg}.


Finally, we note that there are other possible realistic scenarios to be explored deviating from continuous random network models. One intriguing direction is the fate of topological states under the occurrence of crystalline regions. For example, amorphous Si can have a substantial fraction of crystal structure \cite{Gibson:2010cd,Treacy:2012bk}. Amorphous monolayer carbon has been observed~\cite{Toh:2020dy} to have intercalated regions with perfect hexagonal order and regions described by a continuous random network~\cite{Kapko:2010fy}. These mixed crystal-amorphous order is captured under the crystallite model~\cite{Wright:2013ce}, which has not been explored in the context of topological matter. The effect of other possible characteristics of amorphous lattices, such as hyperuniformity~\cite{Florescu:2009ev}, also remain to be fully addressed in the context of topological phases despite encouraging first progress\cite{Mitchell2018}.

\subsection{Physical properties and theoretical characterization of amorphous topological matter\label{sec:charact}}

One of the difficulties that underlies amorphous topological matter is to signal its existence.
Theoretically, diagnosing topological phases in the absence of crystalline symmetry is well developed due to the previously existing work on the survival of topology in the presence of disorder. Many of the tools developed in such context can be translated to amorphous matter, but, as we shall describe shortly, they also have their limitations. 

\subsubsection{Local markers}

To address the robustness of topological phases to disorder some of the topological invariants that characterize them have been expressed in real space. These topological markers have made their way into aperiodic system, and have been proven useful to characterize amorphous matter as well~\cite{Agarwala:2017jv,Mitchell2018,Bourne:2018jr}.

Chern insulators are the simplest to diagnose. As we discussed above, Eq.~\eqref{eq:sigmad} determines the Hall conductivity through an integral of the Berry curvature. Experimentally, a quantized Hall conductivity\cite{Haldane04}, or a quantized circular dichroism~\cite{Tran2017} will signal this type of topological insulator, and can be measured independent of periodicity. Theoretically, in an amorphous structure we cannot rely on the momentum space formulation of Eq.~\eqref{eq:sigmaxy}, and several alternatives have been developed~\cite{Kitaev:2005ik,Prodan:2010jd,Bianco2011,Bourne:2018jr}. Most of them can be obtained by Fourier transforming Eq.~\eqref{eq:sigmaxy} into real space, leading to an expression in terms of the projector onto the occupied ($\hat{P}$) and unoccupied ($\hat{Q}=1-\hat{P}$) bands such that
\begin{equation}
    C = -2\pi i  \ \sum_{\text{all } \textbf{r}_{i}} \bra{\textbf{r}_{i}} \big[ \hat{Q}\hat{x},\hat{P}\hat{y} \big] \ket{\textbf{r}_{i}} \equiv  \sum_{\text{all } \textbf{r}_{i}}C_{xy}(\textbf{r}_{i}).
    \label{CDLCM}
\end{equation}
Where $C_{xy}(\textbf{r}_i)$ is a local quantity known as the local Chern marker\cite{Bianco2011}. 
Interestingly, the local Chern marker can be written in terms of the single-particle density matrix $\rho_{n}(\textbf{r}_{i},\textbf{r}_{j}) = \left.\bra{\textbf{r}_{i}}{n}\right\rangle\left.\bra{n}{\textbf{r}}_{j}\right\rangle$
as~\cite{Kitaev:2005ik,Irsigler19} 

\begin{equation}
\label{eq:LCM}
    C_{xy}(\textbf{r}_{i}) = -4\pi\sum_{\textbf{r}_{j},\textbf{r}_{k}} \sum_{\substack{E_{l}<0 \\ E_{m}>0 \\ E_{n}<0 }}\text{Im}\left[ \rho_{l}(\textbf{r}_{i},\textbf{r}_{j})\rho_{m}(\textbf{r}_{j},\textbf{r}_{k})\rho_{n}(\textbf{r}_{k},\textbf{r}_{i}) \right] x_{j}y_{k} \  .
\end{equation}
If any pair of $\mathbf{r}_i$, $\mathbf{r}_j$, $\mathbf{r}_k$ is equal, the quantity inside the square brackets is purely real and the local Chern marker vanishes. The Chern number is obtained by summing over all possible triangles of sites, and a direct computation shows that triangles formed by nearest-neighboring and next-nearest-neighboring sites dominates the sum~\cite{Irsigler19}. In this sense, even though the topological information is non-local, the local Chern marker is a quasilocal indicator of topology because it is to leading order given by a sum of local triangles~\cite{Mitchell2018,Irsigler19}.

The trace of the local Chern marker results in the Chern number upon summing over bulk sites, but it always vanishes exactly for finite systems. Notice that the local Chern marker is a trace over a commutator defined in a finite Hilbert space, which necessarily vanishes. Mathematically this is reasonable, since topologically, the manifold of a system with open boundary conditions is flat and trivial, unlike a periodic system, which has the topology of a torus. In Fig.~\ref{fig:amorphous models}\textbf{b} we show the local Chern marker for the topological Weaire-Thorpe Chern insulator model with parameters within the topological phase. The bulk is quantized to the Chern number of the periodic system ($C=-1$), while the boundary contributes with a large contribution of opposite sign so that the average over the whole system remains zero. In a trivial phase we would have observed a vanishing Chern number both in the bulk and at the edge.

Interestingly there is no clear relation between the original local Chern marker formula used by Kitaev~\cite{Kitaev:2005ik}, essentially equivalent to Eq.~\eqref{CDLCM}, and the Kubo formula written in real space that determines the Hall conductivity~\cite{Bourne:2018jr}. They both seem to capture well the presence of a Hall insulator, and the survival of quantization in amorphous lattices can be proven rigorously, including spectral and mobility gaps~\cite{Bourne:2018jr}. 

Beyond these mathematical subtleties, physically the trace over the local Chern marker can be directly related to the frequency integrated circular dichroism, the total differential absorption between left and right circularly polarized light. This connection can be traced to Bennett and Stern\cite{Bennett:1965jy}, which observed that circular dichroism is determined by the Hall conductivity, later written as a trace over the local Chern marker\cite{Souza2008,Tran2017}. Therefore, a finite Chern number implies a quantized circular dichroism, an observation that extends to amorphous systems~\cite{Marsal20}.

The Bott index~\cite{Loring:2010jh} is an alternative topological marker to the local Chern marker. A finite Bott index signals the absence of a localized basis to represent the state.  It avoids the subtleties of defining the position operators $\hat{x},\hat{y}$ in periodic systems since it is defined through exponentials of position operators. Explicitly 
\begin{equation}
\label{eq:Bott}
 B = \dfrac{1}{2\pi}\mathrm{Im}\left\lbrace\mathrm{Tr}[\log(WUW^{\dagger}U^{\dagger})]\right\rbrace,
\end{equation}
where $U=\hat{P}\mathrm{exp}(i\Theta)\hat{P}$, $W=\hat{P}\mathrm{exp}(i\Phi)\hat{P}$ and $\Theta$ and $\Phi$ are diagonal matrices that encode the position operators $\hat{x}$ and $\hat{y}$ re-scaled to the interval $[0,2\pi)$.
The Bott index captures amorphous Chern insulator phases~\cite{Agarwala:2017jv} yet it is not straightforwardly connected to the local Chern marker, since no local version of the Bott index exists.

%
%

Topological markers are well developed for Chern insulators, but their generalization to other topological phases is challenging. As we reviewed in Section \ref{sec:strongTI} there are many classes of topological insulators, depending on the symmetries that protect them. For $\mathbb{Z}_2$ topological insulators in two-dimensions with conserved spin projection in the $s_z$ direction, a spin-resolved version of the local Chern marker can be constructed~\cite{Amaricci17}. In more general situations spin-conservation is broken, but the $\mathbb{Z}_2$ topological insulator remains, and in this case, no local invariant exists. For three-dimensional $\mathbb{Z}_2$ topological insulators it is possible to define a three-dimensional topological polarization $\theta$, that, as was mentioned in Section~\ref{sec:strongTI}, should be quantized to $\theta = 0 ,\pi$. This quantity shows similar features to the one-dimensional polarization. It is only defined modulo $2\pi$ which causes troubles when formulating a closed real-space expression~\cite{Malashevich:2010hn,Coh:2011gq,Olsen:2017bz,Rauch:2018gz}. A formula close to Kitaev's original local Chern marker proposal was developed recently~\cite{LiMong19}. 

An interesting case where invariants can be efficiently computed is the case of mirror-topological insulators~\cite{Varjas20}. The eigenvalues can be labelled by their mirror eigenvalue, allowing for their local Chern marker to be computed. These markers are promising options yet are still to be applied to amorphous systems. Additionally, the extension of local Chern markers to other topological phases, like superconductors is still to be developed.

\subsubsection{Responses to external fields \label{sec:Responsesam}}

As we introduced in Sections \ref{sec:basicquant} and \ref{sec:strongTI} topological phases are defined by quantized observables. By calculating or measuring these, it is possible to signal topology. A particularly well known example in two dimensions is that of the Hall conductivity which is quantized to integers in units of fundamental constants. In three dimensions a notable example occurs in $\mathbb{Z}_2$ topological insulators, which host a topological magnetoelectric effect~\cite{Qi:2008eu,Essin:2009kb}.

The topological magnetoelectric effect is a rather unique response of three-dimensional $\mathbb{Z}_2$ topological insulators to external electromagnetic fields. We are accustomed to describing a medium by its dielectric permeability and magnetic permittivity. These enter the definitions of the displacement field $D_i = \varepsilon_{ij}E_j$ and the induction $H_i =\mu^{-1}_{ij}B_j$ in terms of the electric $E_j$ and magnetic $B_j$ fields respectively. However, in magneto-electric materials $D_i$ and $H_i$ are allowed to depend on $B_i$ and $E_i$ respectively. To lowest order in the fields magnetoelectricity can be captured by a term $\alpha_{ij}E_iB_j$ in the effective free energy of the material. The tensor $\alpha_{ij}$ is the magnetoelectric tensor
\begin{equation}
    \alpha_{ij} = \left.\dfrac{\partial P_j}{\partial B_i}\right\vert_{\mathbf{E}=0} = \left.\dfrac{\partial M_j}{\partial {E}_i}\right\vert_{\mathbf{B}=0}.
\end{equation}
It accounts for the change in polarization $P_j$ or magnetization $M_j$ caused by a magnetic $B_i$ or an electric field $E_i$, respectively. Soon after three-dimensional $\mathbb{Z}_2$ topological insulators were predicted it was realized that there is an isotropic contribution to $\alpha_{ij}$  determined by the topological invariant introduced in \ref{sec:strongTI} for this class~\cite{Qi:2008eu,Essin:2009kb}, $\theta=0,\pi$, mod $2\pi$. Specifically, the isotropic part of the magnetoelectric tensor is topological and defined by $\alpha^{\mathrm{iso}}_{ij}= \theta (e^2/2\pi h) \delta_{ij}$~\cite{Malashevich:2010hn,Coh:2011gq,Olsen:2017bz,Rauch:2018gz}.

There are several physical consequences that can be traced back to $\alpha^{\mathrm{iso}}_{ij}$. They include a surface Hall effect~\cite{Essin:2009kb,Rauch:2018gz} but also more exotic responses such as a quantized Kerr and Faraday rotation~\cite{Wu1124}, a repulsive Casimir effect\cite{Grushin2011a} and the Witten effect\cite{Rosenberg:2010gc}. While all of them can signal how this topological state emerges even in the absence of translational invariance, only the Witten effect has been used so far as a theoretical diagnostic of putative amorphous $\mathbb{Z}_2$ topological insulators~\cite{Mukati2020}. The Witten effects predicts that if a magnetic monopole is to be placed inside this type of topological insulator, it will bind a half integer electric charge forming a composite particle known as a dyon, enabled by the magnetoelectric coupling $\alpha^{\mathrm{iso}}_{ij}$. By monitoring this charge as a function of model parameters it is possible to map a topological phase diagram~\cite{Mukati2020}. 

Since the notion of topological marker has not been developed for all symmetry classes, the explicit calculation of topological responses can serve as a theoretical diagnostic of topological states. They have the benefit to generate specific physical predictions that can be used to discover topological phases, as we will discuss in Section \ref{sec:expam}.

\subsubsection{Symmetry indicators and effective Hamiltonians}

Although calculating the local Chern markers is possible, and calculating physical responses is necessary, these are not always computationally efficient to diagnose and classify topology. Developing a classification that can allow a systematic prediction of new materials requires a simple diagnosis of topological phases, which is currently lacking for amorphous systems. Two recent promising directions have emerged: the use of emergent symmetries and the definition of effective Hamiltonians. 

Amorphous matter does not enjoy exact symmetries in general, with the exception of time-reversal in non-magnetic systems, and particle-hole symmetry in superconductors. However it may enjoy average symmetries, that can possibly be exploited. Additionally, even if the whole spectrum is not fully symmetric, special states can be exact eigenstates of a given symmetry. Is it possible to follow these to track how topology changes with different parameters?
The answer is yes, at least for the Weaire-Thorpe models introduced in Eq.~\eqref{eq:Topo_WT}. By construction the sites $i$ are equivalent: upon permutation of orbitals $j$ the Hamiltonian remains invariant. This endows them with a mathematical structure that can be used to predict the evolution of spectral gaps as the parameters are varied~\cite{Schwartz72}. The band edges remain exact eigenvalues of permutation of orbitals, and they can also be assigned a bonding or antibonding character. For a coordination $z$ lattice, there are $2z$ band edges, which can be tracked as a function of parameters by solving exact inequalities~\cite{Marsal20}, monitoring when gap closings occur. 

For the specific case of a threefold coordinated ($z=3$) Weaire-Thorpe-Chern insulator model, the phase diagram can be reproduced by constructing a symmetry indicator that tracks the symmetry character of the band edges. Although there is for the moment no rigorous derivation for this approach, it can be justified by a parallelism to the theory of symmetry indicators in crystals~\cite{Kruthoff17,Po:2017ci,Po20,Vergniory:2019ub,Zhang:2019tp,Tang18}. These are based on an idea that has been central to our discussion: in order to classify topological crystals it is paramount to define what trivial means. In Section \ref{sec:basicquant} we defined the atomic insulator, for which the Wannier states are localized at atomic sites and which is for our purposes trivial. By defining the trivial limit, we defined a reference by which insulators can be compared, diagnosing topology through gap closings.

In the Weaire-Thorpe-Chern insulator there are two trivial limits $W=0$ and $V=0$, of decoupled sites and decoupled dimers respectively. Taking the $W=0$ as the reference trivial insulator we can construct the following indicator
\begin{equation}
\label{eq:symind}
    C = \sum_{n\in \mathrm{filled}} m_n - \sum_{n\in \mathrm{filled}}m^{W=0}_n \hspace{0.2cm} \mathrm{mod}\hspace{0.1cm}3
\end{equation}
where $m=0,\pm 1$ label the band edges with eigenstates of the permutation symmetry with eigenvalue $\xi_m = e^{i 2\pi m/z}$. The first term tracks band edge inversions, or in other words, when the gaps close. The second term acts as a reference for the trivial state.
This formula exactly captures the phase diagram of the threefold coordinated ($z=3$) Weaire-Thorpe-Chern insulator model, as shown in Fig.~\ref{fig:amorphous models}\textbf{b}. 

Another promising route to signal topology in amorphous systems is the use of an effective Hamiltonian~\cite{Varjas2019}. The central idea is to project the real space Hamiltonian into a basis of plane waves with well defined momentum $\mathbf{k}$, with the expectation that the topological information survives along the way.  The effective Hamiltonian is defined as $H_{\mathrm{eff}}(\mathbf{k}) = G_{\mathrm{eff}}(\mathbf{k})^{-1} + E_F$ by projecting the single-particle Green's function onto plane-wave states
\begin{equation}
\label{eqn:Geff}
G_{\rm{eff}} (\mathbf{k})  = \bra{\mathbf{k}} G \ket{\mathbf{k}},
\end{equation}
using the Green's function $G = \lim_{\eta\to 0} \left( H - E_F + i \eta\right)^{-1}$ of the full Hamiltonian with $E_F$ chosen to be in a gap. The key property of $H_{\mathrm{eff}}(\mathbf{k})$ is that its gap closes only when the gap of the full Hamiltonian closes~\cite{Varjas2019}, thus retaining information about the topological state it describes. 

The effective Hamiltonian approach has been successful to track gap closings in topological phases in quasicrystals~\cite{Varjas2019} and applied to the topological Weaire-Thorpe-Chern model~\cite{Marsal20}. However, in the latter case $H_{\mathrm{eff}}(\mathbf{k})$ does not recover the full phase diagram due to numerical instabilities, more pronounced in regions with a small gap. Should these problems be resolved, $H_{\mathrm{eff}}(\mathbf{k})$ can constitute a promising tool to generalize the symmetry classifications of topological insulators to amorphous systems.

\subsubsection{Other probes}

We conclude this section by mentioning a few additional properties that remain to be explored in the context of topological amorphous matter.
Recently it was suggested that amorphous topological lattices may display distinct phase transitions compared to the crystalline counterparts. Strikingly, the critical exponent that determines how the correlation length $\xi$ diverges as an amorphous Hall insulators consisting of growing domains reaches the percolation transition, was reported to be different than previously found for two dimensional electron gases in a magnetic field and Chern insulators~\cite{Sahlberg20,Ivaki2020}. Further numerical studies will be helpful to establish solid evidence of this difference, which could be in principle addressed in experiments.

Lastly, other approaches developed for disordered systems could be also important in the development and signaling of amorphous phases. The level statistics of the entanglement spectrum~\cite{Prodan:2010eq}, or scattering matrix approaches~\cite{Fulga12} serve as alternatives to local markers numerically signal topological phases.

\subsection{Experimental realizations and promising platforms \label{sec:expam}}

Translational invariance is heavily used to interpret experiments and identify topological states. A prime example is angle-resolved photoemission spectroscopy (ARPES), which generated the most explicit images of the energy-momentum conical dispersion of the surface Dirac cones of three-dimensional topological insulators, convincing the community that these topological insulators existed~\cite{Hasan:2010kua,Qi2011}. 
For amorphous systems, the absence of momentum as a good quantum number suggests that typical tools sensible to topological surface states, including ARPES, magnetic quantum oscillations, quasiparticle interference, scanning tunneling microscopy, transport phenomena, or optical responses should be revisited to scrutinize the fingerprints of amorphous topological states. Some encouraging progress suggests that this endeavour is not hopeless.

\subsubsection{Condensed matter systems}

Crystalline Bi$_2$Se$_3$ has been, since very early on, the paradigmatic example of a topological insulator\cite{Zhang:2009ks}. It has a bulk gap of around $0.3$eV, with a single two-dimensional Dirac surface state. The chemical potential of most samples lies above the gap, turning this material into a metal. This fact is not a drawback for  ARPES, which has been used to image the surface Dirac cone perfectly, but it is a drawback for applications which aim to exploit the robustness of its surface Dirac cone. 

That topological properties can survive in non-crystalline solids was hinted by early experiments. For example sputtered Bi$_x$Se$_{1-x}$\cite{DC:2018uf} have been shown to  present weak anti-localization effects, discussed below, associated with spin-momentum locking of a putative Dirac surface state\cite{Sahu:2018cm}. Additionally granular Bi$_2$Se$_3$  was shown to host topological surface states similar as those associated to the bulk crystal\cite{Banerjee:2017jd}.

In late 2019 amorphous Bi$_2$Se$_3$ was studied using transport, ARPES and spin-resolved ARPES, showing signatures consistent with an amorphous topological insulator phase in the solid-state\cite{Corbae:2019tg}. High resolution transmission electron microscopy showed a diffraction pattern typical of amorphous matter, with well defined coordination and inter-atomic distance, similar to the crystalline case, signaled by a diffuse but well defined ring corresponding to the short-range ordering of nearest neighbors. The data showed no signs of nanocrystals or lattice fringes, consistent with the amorphous nature of the samples. As anticipated in Section \ref{sec:modelsamorp} these observations suggest that Weaire-Thorpe type models, which fix coordination, are well suited to describe amorphous topological insulators.

As a function of temperature $T$, the $e^{-(T_0/T)^{1/4}}$ behaviour of the conductivity observed in these samples is consistent with that expected from a variable range hopping model~\cite{Hill76}, indicating that the bulk is largely Anderson localized. At low temperatures, when time reversal symmetry is present the spin-momentum locking of putative Dirac surface states suppresses back-scattering. The first quantum correction to the conductivity is positive and increases conductivity, a phenomenon that is known as weak anti-localization, which supports the robustness of the Dirac surface states (and other Dirac metals like graphene) to disorder\cite{Bardarson:2013cn}. When a magnetic field is applied, time-reversal symmetry is broken and backscattering increases leading to a positive magnetoresistance. This phenomena has been observed in crystal topological insulators, and taken as strong indication of the presence of surface states\cite{Brahlek:2015eh,Hasan:2010kua}. Upon applying a magnetic field to amorphous Bi$_2$Se$_3$ the sample showed weak anti-localization at small magnetic fields. The Hikami-Larkin-Nagaoka formula for weak anti-localization\cite{Hikami80} indicated that the magnetoconductance can be well explained by two conducting channels at low temperatures. This is consistent with two surface conducting states and an Anderson localized bulk\cite{Corbae:2019tg}. 

\begin{figure}
    \centering
    \includegraphics[width=\linewidth]{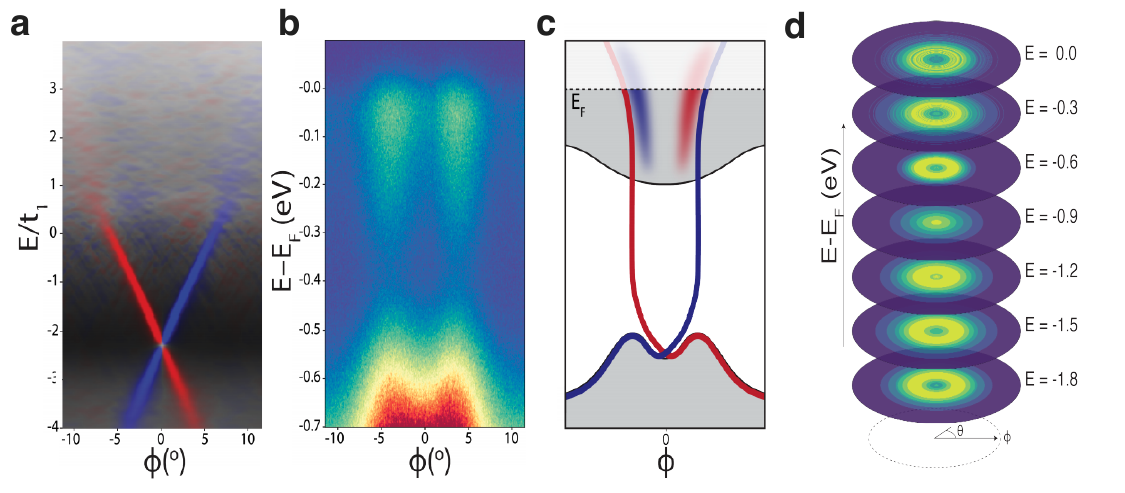}
    \caption{\textbf{ARPES on Amorphous Bi$_2$Se$_3$} \textbf{a} An amorphous tight-binding model constructed with hoppings based on the model that describes crytalline Bi$_2$Se$_3$ shows a Dirac surface state as a function of detector angle $\phi$. The arrows indicate the opposite spin projections of the two surface state branches. \textbf{b} and \textbf{d} show ARPES data at different photon energies as a function of the detection angles $\phi$ and $\theta$. They evidence the presence of a spectral gap and crossing point, consistent with a surface Dirac cone. \textbf{c} The spin-ARPES results (not shown) support a schematic dispersion relation where spin projections reverse in the bulk bands compared to the in-gap surface state, resembiling that observed in crystalline Bi$_2$Se$_3$ (adapted from Ref.\cite{Corbae:2019tg}).}
    \label{fig:amorphousexp}
\end{figure}

Perhaps the most striking observation in this sample was that of a surface Dirac cone in ARPES and spin-resolved ARPES\cite{Corbae:2019tg}. 
Although momentum is no longer a good quantum number, it is possible to represent the energy dispersion as a function of emission angle $\phi$ (see Fig.~\ref{fig:amorphousexp}\textbf{b}).
The spectrum shows two nearly vertical features, consistent with a putative surface state, that merge with a high intensity region, consistent with a bulk valence band.
By angle integrating the ARPES signal the band gap was estimated to be close to $0.3$ eV. The spin-resolved ARPES supports the topological state interpretation. Resembling a striking similarity with crystal Bi$_2$Se$_3$~\cite{Jozwiak:2016jv}, the experiment shows a clear spin asymmetry of the bulk and surface states. The putative surface band spin asymmetry is consistent with spin-momentum locking expected from Eq.~\eqref{eq:Dirac}.

Although it might be striking that a spin polarized cone could be seen in photoemission, theoretical models support this picture. By calculating the spectral function of a model describing hoppings that lead a $\mathbb{Z}_2$ topological insulator on an amorphous lattice\cite{Agarwala:2017jv}, with fixed coordination~\cite{Corbae:2019tg}, a Dirac cone emerges with well defined spin-structure (see Fig.~\ref{fig:amorphousexp}\textbf{a}). It arises on top of blurry bulk states, defying the intuition that amorphous materials should only show only a noisy ARPES spectrum.

It is likely that other amorphous condensed matter systems can join the list of physical systems that present topological properties. Recent experiments observed that charge to spin conversion of sputtered amorphous Bi$_x$Se$_{x-1}$, consistent with the survival of topological properties~\cite{DC:2019dj,Sahu19}. Additional, recent theoretical predictions offer alternatives to realize topological states in Shiba glasses emergent from randomly distributed magnetic impurities on the surface of a Rashba superconductor, which could realize amorphous topological superconductivity~\cite{Poyhonen2017}.  However, the difficulty in establishing the requirements to discover amorphous topological solids still precludes the task to predict and identify them in experiment. This impediment is not shared by synthetic systems, which we discuss to conclude.

\subsubsection{Synthetic systems}

The first and currently the only experimental demonstration of topological amorphous systems in synthetic matter was realized in arrays of coupled gyroscopes~\cite{Mitchell2018}, soon after the original proposal that amorphous states could exist~\cite{Agarwala:2017jv}. Pairs of mechanical objects, in particular coupled gyroscopes, can be engineered to have non-reciprocal couplings, that effectively break time-reversal symmetry and allow for a chiral propagating mode at their boundary. This was achieved in a human size experiments at ETH Zurich where coupled pendula where designed to host a topological mode that propagated chiraly at the boundary, in analogy to a Chern insulator. Using a similar idea, Mitchell et al.\cite{Mitchell2018} carried out the same principle on a significant number of amorphous lattices (see Section \ref{sec:modelsamorp} for their description) of coupled gyroscopes. Their findings demonstrated experimentally that amorphous topological states can become an experimental reality for the first time.

Photonic lattices are another promising system to realize topological amorphous states~\cite{Rechtsman:2011ej}. Beyond the amorphous photonic Chern insulator proposal discussed above \cite{Mansha:2017be}, it is encouraging that the Weaire-Thorpe model has been recently realized in photonic topological lattices, achieving significant photonic band gaps~\cite{Florescu:2009ev}. Combining this realization with a periodically modulated waveguide in a third perpendicular direction effectively breaks time-reversal symmetry and acts like a magnetic field, experimentally realized in~\cite{Rechtsman:2013fe}. As for quasicrystals~\cite{Bandres:2016gx} this is a very promising route to realize the topological Weaire-Thorpe model proposed in Ref.~\cite{Marsal20}.

It is reasonable to expect that the same synthetic platforms that have been successful in implementing topological band structures can serve to realize analogues of solid-state topological amorphous states~\cite{Bourne:2018jr}.
Specifically, in addition to those mentioned above, acoustic systems, topo-electrical circuits, ultra-cold atoms trapped in optical lattices and mechanical systems are all promising platforms to experimentally explore the properties of amorphous topological states.

\section{Conclusions and outlook}

The potential of amorphous topological matter seems extraordinary. The lack of translational invariance is motivating to push the limits of our descriptions of topological matter, and how to identify it in experiment. The ubiquity of amorphous solids in technology adds even more motivation to propose new device principles, that combine the robustness of topology and the versatility of amorphousness.  

The quest for a complete classification of topological amorphous solids that can aid discovery is in its infancy. Achieving this milestone can aid to discover topological insulators beyond the only known amorphous topological insulator in the solid state, amorphous Bi$_2$Se$_3$~\cite{Corbae:2019tg}. It can also aid to find amorphous topological matter that surpasses their crystal counterparts. For example Bismuth is superconducting at 6K in its amorphous form, a $10^4$ increase from its crystalline counterpart. Extending this enhancement to amorphous superconducting materials with strong spin-orbit coupling can be the door to robust topological superconductivity. 

Even more boldly, it is appealing to search for new topological phases that can only be realized in amorphous lattices. We have witnessed similar developments in quasicrystals~\cite{Varjas2019}, where an eight-fold rotational symmetry, absent in crystals, protects a higher order topological phase. 

Many other questions remain to be addressed, such as the effect of interactions,
the interplay between topological (or lattice) disorder and site disorder, or the nature of different topological phase transitions~\cite{Sahlberg20}. Together with the pressing need to discover new physical systems in both condensed and synthetic matter, it is hardly doubtful that this new field will thrive further, combining the efforts of two disconnected communities, those studying topological and amorphous states of matter.

\section{Acknowledgements}

The author is indebted to A. Cano, P. Corbae, S. Ciocys, E. Dresselhaus,  Q. Marsal,  O. Pozo, R. Queiroz, C. Repellin, B. Sbierski, and D. Varjas for stimulating discussions and related collaborations. The author especially thanks J. H. Bardarson and M. A. Ramos for the critical reading of this chapter, and A. Cano for discussions on its structure. The author acknowledges financial support by the ANR under the grant ANR-18-CE30-0001-01 (TOPODRIVE) and the European Union Horizon 2020 research and innovation programme under grant agreement No 829044 (SCHINES).

\bibliographystyle{ws-rv-van}

\end{document}